\documentclass[%
 reprint,
 nofootinbib,
 amsmath,amssymb,
 aps,
]{revtex4-2}

\usepackage{graphicx}
\usepackage{appendix}
\usepackage{dcolumn}
\usepackage{bm}
\usepackage{hyperref}
\usepackage{xcolor}
\usepackage{siunitx}
\usepackage{mhchem}
\usepackage[capitalize]{cleveref}
\usepackage{natbib}  
\DeclareSIUnit\Molar{\textsc{M}}
\begin{document}

\title{Viscophoretic particle transport}

\author{Vahid Khandan,\textsuperscript{1}  Vincent J. P. Boerkamp,\textsuperscript{2} Abbas Jabermoradi,\textsuperscript{2} Mattia Fontana,\textsuperscript{2} Johannes Hohlbein,\textsuperscript{2,3} Elisabeth Verpoorte,\textsuperscript{1} Ryan C. Chiechi,\textsuperscript{4,}* and Klaus Mathwig\textsuperscript{1,5,}}
\email{ryan.chiechi@ncsu.edu}
\email{klaus.mathwig@imec.nl.}

\affiliation{
\textsuperscript{1}University of Groningen, Groningen Research Institute of Pharmacy, Pharmaceutical Analysis, P.O. Box 196, XB20, 9700 AD Groningen, the Netherlands\\
\textsuperscript{2}Laboratory of Biophysics, Wageningen University \& Research, Stippeneng 4, 6708 WE Wageningen, the Netherlands\\
\textsuperscript{3}Microspectroscopy Research Facility, Wageningen University \& Research, Stippeneng 4, 6708 WE Wageningen, the Netherlands\\
\textsuperscript{4}Department of Chemistry \& Organic and Carbon Electronics Laboratory, North Carolina State University, Raleigh, NC, 27695 USA\\
\textsuperscript{5}imec within OnePlanet Research Center, Bronland 10, 6708 WH Wageningen, the Netherlands
}

\date{\today}

\begin{abstract}
Viscosity is a fundamental property of liquids and determines the diffusivity of suspended particles. A gradient in viscosity leads to a gradient in diffusivity, yet it is unknown whether such a gradient can lead to \emph{directed} transport of particles.  In  this work, we generate a steep, stable viscosity gradient in a microfluidic channel and image the resulting transport of suspended nanoparticles at the single-particle level using high-resolution microscopy. We observe high \emph{viscophoretic} drift velocities that significantly exceed theoretical predictions. In addition, we utilize viscophoresis for a new type of particle trap. We provide a first quantification of a transport phenomenon that is of importance in any system and any application exhibiting viscosity gradients, for example in separation using membrane technology as well as in inter- and intracellular biomolecular transport.
\end{abstract}

\maketitle


\
Transport phenomena of particles and molecules in solution due to gravity, electrical fields, and concentration gradients have been studied in great detail and are well understood \citep{Kynch1952,Tiselius1937,Fick1855,Feynman2010}. 
In stark contrast, the transport of particles in a gradient of viscosity has yet to be investigated systematically at the single-particle level. 
Viscosity is the property of liquids that determines the transport of solvent, particles, molecules, and ions, it is fundamental to the ubiquitous role of liquids in living and artificial systems \citep{Vand1948,Homsy1987,Shinitzky1976,Bera2022}. 
Although often treated as a parameter of a bulk liquid, a \textit{gradient} in viscosity generates a spatial gradient in the diffusion coefficient that drives transport of molecules and particles, namely \emph{viscophoresis}. 
Such a phenomenon would play an important role at interfaces and in small, confined spaces such as cellular compartments and microfluidic channels (both natural and artificial), yet it is not understood how a gradient in viscosity causes the anisotropic motion of particles and molecules. Despite this ubiquity, and in contrast to other short-range diffusional transport effects such as thermodiffusion \citep{Duhr2006}, diffusiophoresis \citep{Abecassis2008}, diffusioosmosis \citep{Williams2020,Shim2022}, Brownian motors \citep{Hanggi2009, Matthias2003} and swimmers \citep{Li2008}, or ballistic Brownian motion \citep{Huang2011, Franosch2011},  viscophoresis has yet received little experimental attention \citep{Wiener2018} because of the considerable experimental challenge of the controlled generation of steep viscosity gradients and subsequent isolation of viscophoretic effects from masking phenomena.

\begin{figure}[b]
\includegraphics[width=8.7cm]{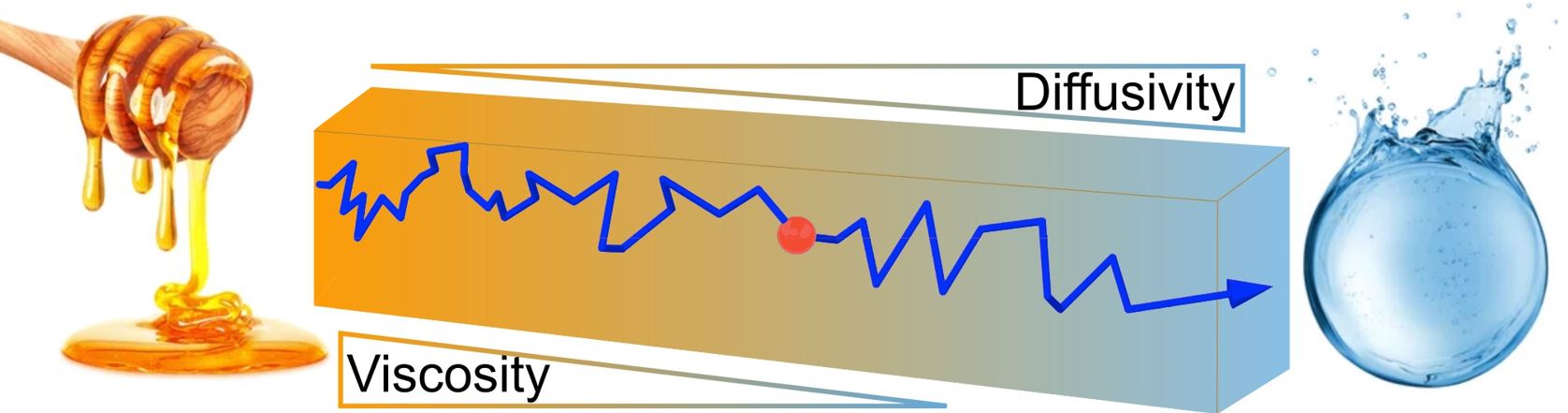}
\caption{\label{fig:viscosity1} \textbf{Schematic of \emph{viscophoretic} transport.} A particle is driven from viscous to a low-viscosity fluid from left to right. Its diffusion coefficient in this direction increases, and steps of the Brownian walk become longer---leading to net transport toward lower viscosity.}
\end{figure}

Thus, solving the problem of measuring viscophoresis is the first step to understanding how and why it is exploited in natural systems such as the microviscosity of cellular organelles \citep{Chambers2018}  and to exploiting phoretic phenomena in artificial systems, for example, for sensing and manipulating molecules and particles. Understanding the effects of viscosity gradients at small length scales will likely provide valuable insights into larger length scales as well, such as magma viscosity gradients in volcanoes \citep{Yamasaki2012} and gradients in planets \citep{Morison2021} or stellar plasma \citep{Bitsch2014}. 
Here, we experimentally answer the fundamental question of how the motion of a particle in solution is affected by a gradient in viscosity through direct observation.

The basic concept of viscophoretic transport is shown in \cref{fig:viscosity1}. A particle or molecule in solution undergoes a diffusive Brownian walk in a gradient of viscosity. As viscosity decreases from left to right, the diffusivity of the particle increases accordingly. As a result, Brownian steps toward the right (higher diffusivity) become larger than toward the left (higher viscosity). Therefore, the particle experiences a net drift towards the right side in the direction of lower viscosity and higher diffusivity. 

We designed microfluidic devices that allow us to generate very steep, stable viscosity gradients (up to $\Delta\eta = 5\,$mPa\,s over 100\,$\mu$m) and that are compatible with high-resolution fluorescence imaging. 
This experimental setup enables the direct observation of drift at the single-particle level by image correlation analysis, demonstrating viscophoresis directly and with different types of particles. Quantitatively, we find that viscophoretic drift velocities are much larger than predicted for nanoparticles of 10\,nm to 100\,nm in diameter.

We use this discovery to demonstrate a new microfluidic principle for trapping and concentrating nanoparticles by combining viscophoretic and diffusiophoretic transport. Compared to the established biophysical toolkit of particle traps \citep{Huang2011}, this new principle promises to uniquely combine
practical advantages: no active feedback control is necessary, and molecules can be trapped under any electrolyte conditions independently of their charge and dielectric properties.

\subsection*{Theory and state of the art}

To determine the drift caused by the viscous gradient in diffusivity, Fick's law 
\begin{equation}
	\bm{J}=-\bm{D}\frac{\partial c(x)}{\partial x}
\label{eq:fick}
\end{equation}
must be generalized for space-dependent diffusivity $\bm{D}=\bm{D}(x)$, where $\bm{J}$ is diffusive flux and $c$ is particle concentration. In such a gradient of diffusivity, 
transport is a phenomenon with multiplicative feedback, i.e., Brownian diffusion exhibits a state-dependent noise intensity \citep{Pesce2013}, and transport cannot be derived from first principles. This ambiguity for evaluating trajectories in non-uniform stochastic noise is called the \textit{It\^{o}-Stratonovich dilemma} \citep{Mannella2012,Kuroiwa2014,Farago2014}. Possible solutions of the noise-induced viscophoretic drift velocity $v_\textrm{VP}$ of a particle include
\begin{equation}
	v_\textrm{VP}=\langle \dot{x}\rangle = \alpha \frac{\textrm{d}  D(x)}{\textrm{d}x} \textrm{ with } \alpha = \{ 0,\textstyle \frac{1}{2},1 \}, 
\label{eq:ito}
\end{equation}
with the  It\^{o} ($\alpha =0$, no viscophoresis),  Stratonovich ($\alpha=\textstyle \frac{1}{2}$) , and isothermal ($\alpha = 1$) choice.

In theoretical work, viscophoretic drift is generally described with $\alpha = 1$, i.e., by the isothermal process that is synonymous with fast drift \citep{Yang2013, ALIZADEH2022545}. To investigate particle transport in viscosity gradients experimentally, Wiener and Stein \citep{Wiener2018} used an H-shaped microchannel geometry and driven flow of carrying solutions with different viscosities to generate the steady-state, non-equilibrium condition of a stable viscosity gradient in a `mixing channel' while suppressing any net mass transport (i.e., flow). In this state-of-the-art research, the authors were then able to infer a viscophoretic drift of ions with $\alpha = 1$ by detecting the ion current through the mixing channel. Instead of performing such an electrical ensemble measurement, we used optical microscopy to image particle-drift---directly---at the level of single particles, and without potential interference of unrelated electrokinetic phenomena.

\subsection*{Experimental concept}
\begin{figure}
\includegraphics[width=8.7cm]{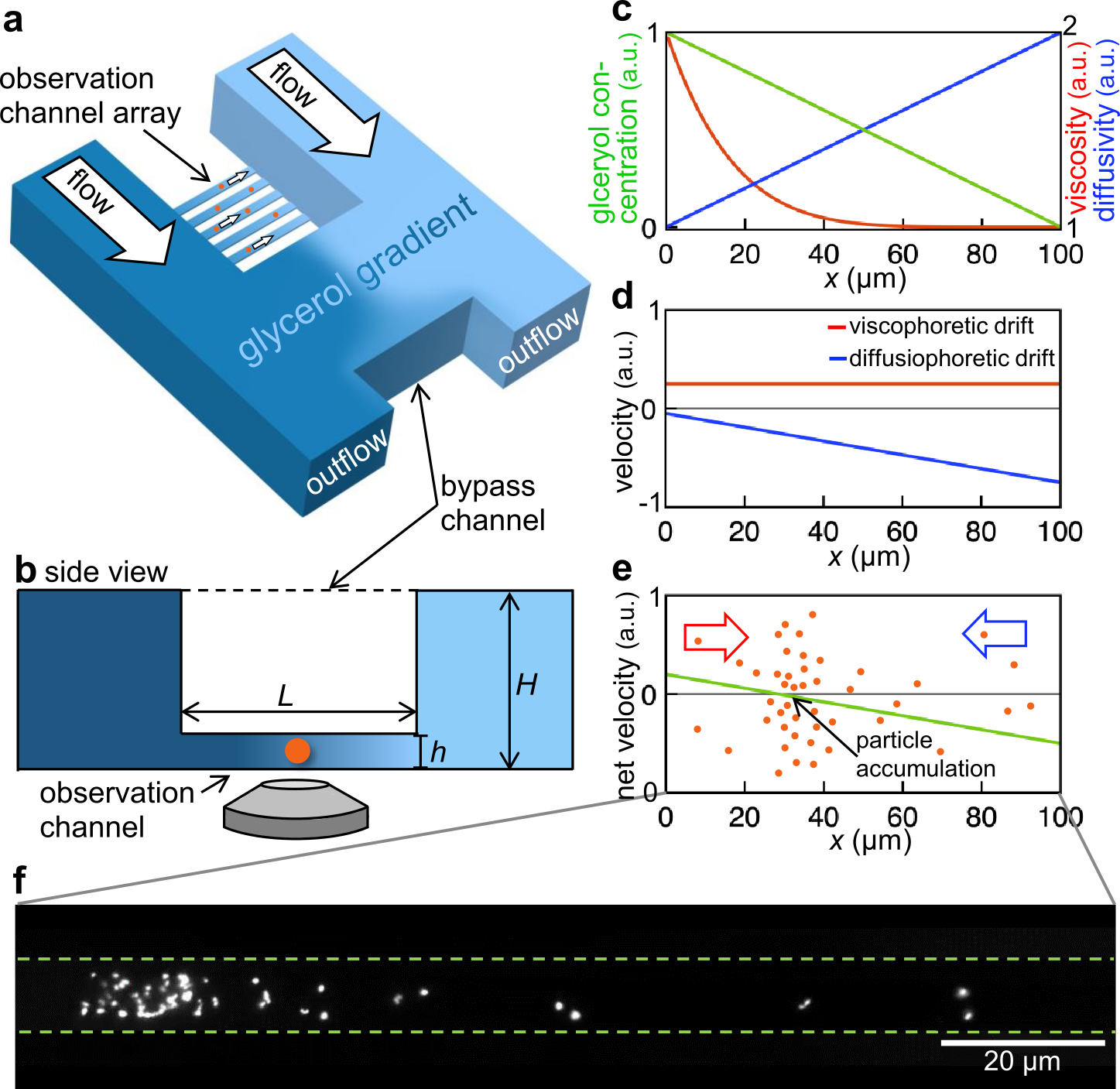}
\caption{\label{fig:concept} \textbf{Experiment and concept of viscophoresis}. \textbf{a}, Top view and \textbf{b}, side view of microfluidic experimental setup. A stable viscosity gradient (blue color gradient) is generated in the 2\,$\mu$m shallow observation channels with suppressed flow using a downstream bypass ($L\sim$ 100\,$\mu$m, $h=$ 2\,$\mu$m, $H=$ 200\,$\mu$m).  \textbf{c}, Gradient of glycerol concentration (green curve) and resulting gradients of viscosity (red) and particle diffusivity (blue) in the observation channel. \textbf{d}, Viscophoretic and diffusiophoretic drift velocities of suspended nanoparticles. \textbf{e}, Resulting overall net velocity leading to particle accumulation in a channel segment. \textbf{f}, Micrograph of 110\,nm-diameter polystyrene nanoparticles accumulating in a 150\,$\mu$m-long channel. Dashed lines indicate microchannel walls.}
\end{figure}

Our experimental setup is shown in \cref{fig:concept}. Using a microfluidic H-channel device fabricated in glass and polydimethylsiloxane (see Methods), two syringe pumps deliver flow of aqueous solutions with different glycerol concentrations, and, thus, different viscosities \citep{Dorset1940}. Flow rates are adjusted to prevent any pressure-driven flow and any solvent mass transport in the observation channel array; and residual flow in the observation channels is suppressed by placing a wide pressure-releasing bypass channel directly downstream of narrow channels with a large hydraulic resistance. These conditions produce a stable viscosity gradient in these observation channels and suppress potential masking phenomena. 
The inner walls of all channels are chemically modified prior to 
experiments to prevent the adsorption of the fluorescent nanoparticles that are introduced at the inlets. We then map nanoparticle-drift in the observation channel by fluorescence microscopy and image correlation analysis (see Methods) \citep{Khandan2024}.  

\Cref{fig:concept}c-e shows the stable gradients that form in the observation channel. The glycerol concentration decreases linearly from left to right, resulting in a viscosity decreasing with $\eta \propto 1/x$ and a linearly increasing nanoparticle diffusivity $D(x)$ \citep{Wiener2018}. Thus, according to \cref{eq:ito} with $\alpha = 1$, a constant positive drift velocity from left to right is expected.

Colloidal particles suspended in a concentration gradient also undergo \textit{diffusiophoresis} \citep{Abecassis2008, Rasmussen2020}. For a nonelectrolyte concentration gradient, particles drift with a diffusiophoretic velocity $v_\textrm{DP}\propto \frac{1}{\eta(x)}\textrm{d}c_\textrm{gly}/\textrm{d}x$ \citep{Anderson1982, Staffeld1989, Paustian2015, Marbach2020}. Here, glycerol molecules are diffusing around polystyrene nanoparticles, causing particle drift toward a higher glycerol concentration from right to left. The absolute value of the velocity $v_\textrm{DP}$ increases linearly from left to right. Serendipitously, diffusiophoretic drift and expected viscophoretic drift have opposite directions, and, thus, the effects can be distinguished directly. The resulting net particle velocity (see \cref{fig:concept}e) decreases linearly along the observation channel starting with a positive velocity dominated by viscophoresis on the left side, crossing over to a negative net velocity (i.e., transport from right to left) dominated by diffusiophoresis  on the right channel side. In effect, particles entering the observation channel from both sides move toward a point at which  $v_{x,\textrm{net}} = v_\textrm{VP} - v_{\textrm{DP}} = $ 0\,$\mu$m, and they accumulate there.

\subsection*{\label{sec:results}Experimental viscophoresis}

\Cref{fig:concept}f shows fluorescent particles with a 110\,nm diameter accumulating in an observation microchannel in which a glycerol viscosity gradient is maintained 
(see \href{https://doi.org/10.6084/m9.figshare.21763559}{https://doi.org/10.6084/m9.figshare.21763559}). 
 Particles enter this channel by diffusion from both inlet channels and are then transported from left to right due to viscophoresis. Opposing diffusiophoretic drift leads to accumulation in the left-most segment of the microchannel. Diffusiophoresis is the dominant transport mechanism along most of the channel, however, as shown in \cref{fig:concept}d, it vanishes though toward the left side, leaving a region over which transport toward the channel center is entirely viscophoretic. Thus, in the absence of viscophoretic transport, diffusiophoretic drift would dominate and no accumulation would occur.

\begin{figure}
\includegraphics[width=8.7cm]{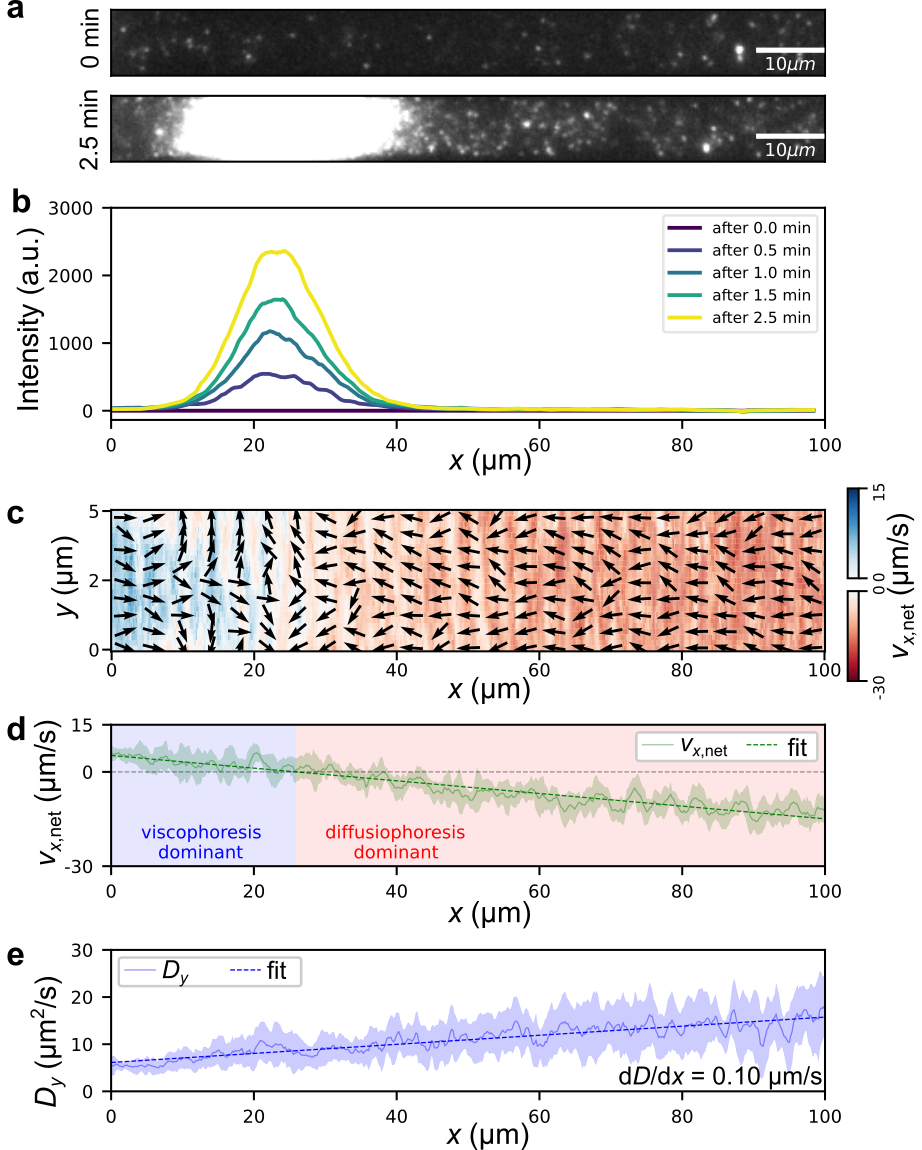}
\caption{\label{fig:results} \textbf{Viscophoretic drift of 28-nm particles.} \textbf{a}, Still images at 0\,min and 2.5\,min and \textbf{b}, fluorescence intensity graph of particles drifting and accumulating over time in an aqueous glycerol gradient ranging from 50\% on the left site to 0\% on the right. \textbf{c}, Map of experimental drift velocity in a microchannel determined via image correlation. Arrows indicate drift direction.  \textbf{d}, Average drift velocity $v_{x,\textrm{net}}$ along the channel in $x$-direction. \textbf{e}, Average diffusion coefficient along the channel (diffusivities are evaluated in vertical $y$-direction). Light green and light blue curves denote 90\% confidence intervals; fits are linear.}
\end{figure}

\Cref{fig:results} shows detailed experimental results demonstrating viscophoresis of particles with a diameter of 28\,nm. 
As particles experience viscophoresis and diffusiophoresis over the length of the observation channels, they continue accumulating in a segment of the microchannel centered where $v_{x,\textrm{net}} = 0\,\mu$m\,s$^{-1}$, which occurs at $x=$ 25\,$\mu$m. \Cref{fig:results}b shows fluorescent intensity profiles of particles accumulating over 2.5\,min. Intensity profiles exhibit the expected Gaussian shape for diffusive broadening of the accumulation zone.   

Fluorescence imaging with high spatial and temporal resolution enables the correlation analysis of video frames to \textit{map} magnitude as well as direction of nanoparticle drift and diffusion in the observation channel. \Cref{fig:results}c shows a map of the time-average \textit{direction} (arrows) of particle drift and its \textit{magnitude} in longitudinal direction ($x$-direction along channel, color map). The overall mean drift  $v_{x,\textrm{net}}$  can  be extracted directly from these data by averaging over the $y$-direction across the channel. The resulting net drift velocity $v_{x,\textrm{net}}$ is shown in \cref{fig:results}d; the light green curve is the confidence interval computed from three separate measurements and the dashed line is a linear fit. Data in \cref{fig:results}a-d clearly show two distinct transport regimes of particle transport in opposite directions dominated by viscophoresis and diffusiophoresis, respectively. In the left-most segment of the channel, diffusiophoresis vanishes and \emph{only} viscophoresis is observed, clearly distinct from diffusive noise, and for 28-nm particles we determined a viscophoretic drift velocity of 5.8\,$\mu$m\,s$^{-1}$ (see below for details). 

To determine particle diffusivities independent of viscophoresis and diffusiophoresis, i.e., without any bias due to drift, we used further image correlation analysis to determine the mean square displacement of particles in $y$-direction, along which drift does not occur. The resulting coefficient $D_y(x)$ is shown in \cref{fig:results}e. This linear diffusivity profile is in agreement with  expectations (blue curve in \cref{fig:concept}c), unambiguously demonstrating the presence of a stable gradient by confirming the absence of any observable nonlinear distortions that would be caused by residual flow.

\subsection*{Excluding interfering transport effects}

To validate our hypothesis that the measured drift velocity is due only to viscophoresis, a new transport phenomenon, we must control for all other possible mechanisms of transport. Electrokinetic effects are not present because there is no applied electric potential; electrostatics are suppressed by a constant concentration of background electrolyte 10\,mM NaOH); and particle-particle and particle-wall interactions are suppressed by surface modification (see Methods). 
The remaining likely source of such parasitic transport is advection in \emph{flow}, i.e., net mass transport of the solvent exerting Stokes forces on the particles under observation. We can eliminate parasitic transport as a possible mechanism for three reasons:

(1) The applied inflow rates are set equal to the inverse of the viscosity ratios to suppress the generation of a pressure difference between inlet channels. In our microfluidic design \citep{Khandan2024b}, the observation channels are very shallow compared to the wide cross-section of the inlet channels, leading to a ratio of the respective hydraulic resistances \citep{Bruus2011} of approximately $1:10^6$.
Consequently, even a large parasitic difference in inlet flow rates of 10\% will  lead to a maximum residual flow velocity of 300\,nm\,s$^{-1}$ in the observation channel, which is negligible compared to the the observed drift velocities and the timescale of particle diffusion and image acquisition ($>$ 3\,$\mu$m in 10\,ms).

(2) We use a bypass channel with a large cross-section positioned directly downstream of the observation channel (see \cref{fig:concept}a,b), which eliminates any residual pressure difference between both ends of the observation channel\citep{Khandan2024b}. 
The direct observation of side-by-side flow of inlet streams in this bypass confirms that pressure difference and cross-flow between the inlet channels is negligible.

(3) Pressure-driven flow would lead to a parabolic profile of Hagen-Poiseuille flow in the channel \citep{Bruus2011}; however, the mapped particle velocity (see \cref{fig:results}c) shows a constant profile in $y$-direction of the cross section in agreement with the visco-/diffusiophoretic transport phenomena described above, confirming the absence of pressure-driven flow.

While microchannel flow can potentially be generated by a chemical (i.e., glycerol) gradient, such diffusioosmosis is negligible in our experiment. 

The only motion that cannot be suppressed is diffusiophoretic because it is a direct consequence of the glycerol concentration gradient. Nonetheless, as described above, the direction of diffusiophoretic motion is opposite that of viscophoretic motion, therefore the accumulation of particles---having eliminated all other effects---can only be explained by the presence of viscophoresis. This newly-demonstrated principle of particle accumulation only occurs in a viscosity gradient.

\subsection*{Quantifying viscophoresis}
\begin{figure*}
\includegraphics[width=17.8cm]{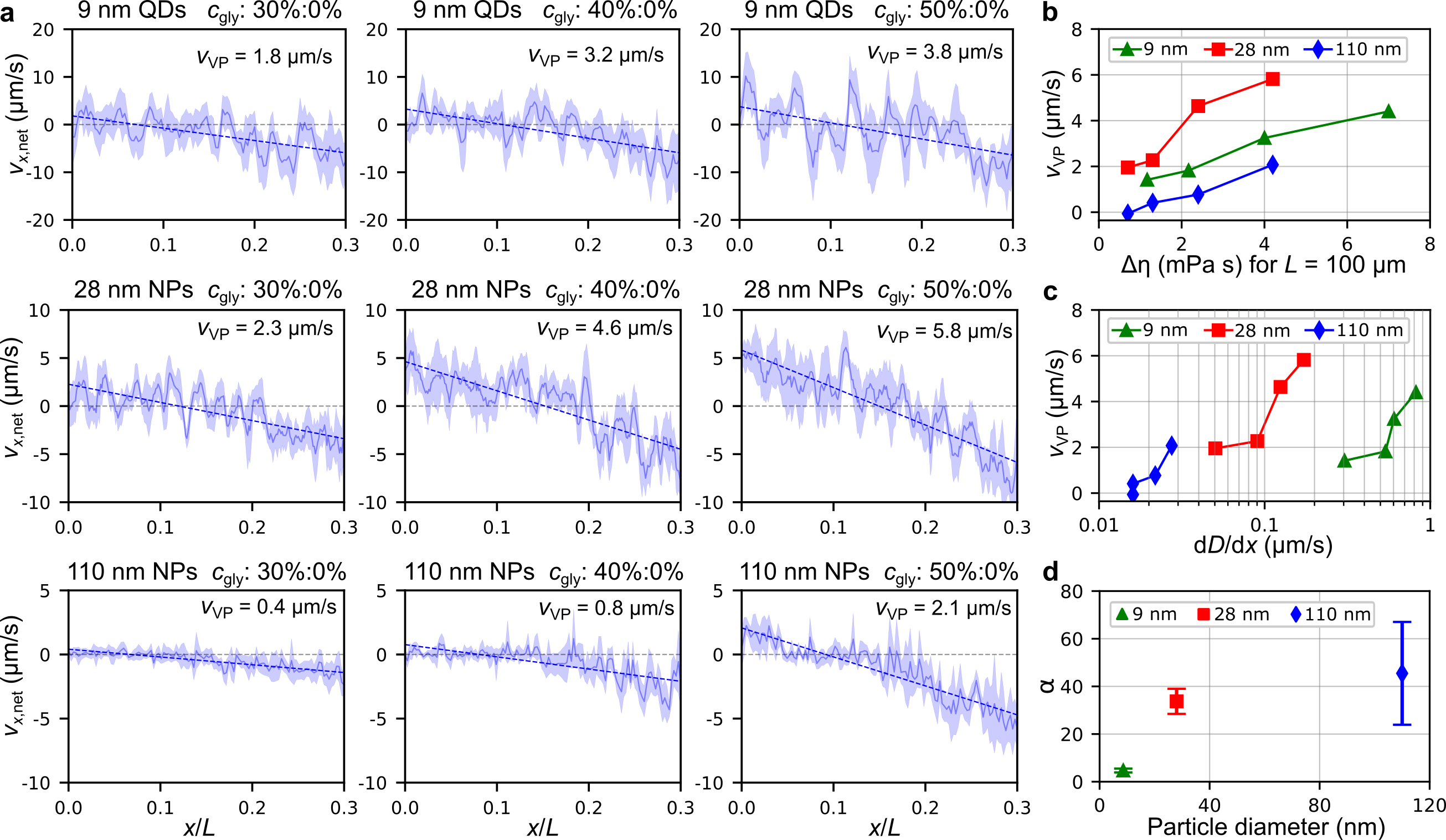}
\caption{\label{fig:wide2} \textbf{Quantifying viscophoresis.} \textbf{a}, Drift velocities $v_{x,\textrm{net}}$ as a function of particle diameter and viscosity gradient. Velocity profiles for 9\,nm, 28\,nm, and 110\,nm particles are shown in rows. Viscosity gradients were varied by varying the weight-percentage of glycerol from 30\,wt\,\% to 50\,wt\,\% at the left inlet (columns). $v_{x,\textrm{net}}$ are determined by averaging mapped velocities in $y$-direction. Data and confidence intervals on the left 30\% of microchannels are shown; fits are linear. \textbf{b}, Viscophoretic drift velocities obtained as $v_\textrm{VP}=v_{x,\textrm{net}} (x=0$\,$\mu$m) of these fits as a function of viscosity difference $\Delta \eta$ of 0\,wt\% glycerol 
 vs. 20\,wt\,\%, 30\,wt\,\%, 40\,wt\,\% and 50\,wt\,\% 
 \textbf{c}, Linear-log plot of $v_\textrm{VP}$ as function of the particle diffusivity gradient d$D/$d$x$. Here d$D/$d$x$ was determined experimentally as shown in \cref{fig:results}e. \textbf{d}, $\alpha$ parameter obtained as slope of a linear fit to curves in panel c, i.e., $v_\textrm{VP}=\alpha\, {\textrm{d}D/\textrm{d}x}$. Error bars show standard error of linear regression. Lines in \textbf{b}, \textbf{c} are guides for the eye.
}
\end{figure*}

In addition to proving the existence of viscophoresis, the ability to observe particle motion directly allows us to quantify the influence of particle size and of the viscosity gradient on viscophoretic drift by varying these parameters experimentally. We observed polystyrene particles with diameters of 28\,nm and 110\,nm and \ce{CdTe} quantum dots with diameter of 9\,nm in different glycerol gradients (see \cref{fig:wide2}). As described above, we determined the magnitude of viscophoretic drift 
at the left channel entrance,
where diffusiophoresis vanishes, i.e.,  $v_\textrm{VP}=v_{x,\textrm{net}} (x= $ 0\,$\mu$m).

The experimental results shown in \cref{fig:wide2}b reveal a positive correlation between the magnitude of viscophoresis and the viscosity gradient. \Cref{fig:wide2}c shows a maximum in drift velocity for 28\,nm particles, however, we were unable to observe a clear dependence of viscophoretic drift velocity on particle size within the accuracy of our measurement. Nonetheless, these results demonstrate the ability of our microfluidic platform to control viscophoretic effects by demonstrating straightforward relationships between viscosity gradients, particle size and viscophoresis.

\Cref{fig:wide2}d plots the parameter $\alpha =  v_\textrm{VP}/\left(\textrm{d}D/\textrm{d}x\right)$ from \cref{eq:ito} as a function of particle size. Surprisingly, these results show a clear positive correlation, ranging from $\alpha = 5$ for particles of 9\,nm in diameter to $\alpha = 45$ for 110\,nm. $\alpha$ clearly increases with particle size. Theory predicts a much slower viscophoretic drift with $\alpha = 1$, i.e., drift velocity is expected to decrease rather than remain steady for larger particles \citep{Yang2013, ALIZADEH2022545}.
Viscophoresis of ions inferred from electrokinetic measurements also determined $\alpha = 1$ \citep{Wiener2018}. That report is not in conflict with our results, since the smallest particles we measured are at least 20 times the diameter of those ions, however, a constant value of $\alpha$ in \cref{eq:ito} clearly does not describe our experimental observations. 

The prediction that viscophoresis varies inversely with particle size makes intuitive sense because diffusivity generally decreases with increasing particle size. We postulate that, while larger particles diffuse more slowly, because the cross-section of a particle increases with its diameter, larger particles sample a larger segment of the viscosity gradient. Therefore, the maximum difference of diffusivity $\Delta D$ at both sides of a particle increases linearly with diameter, leading to viscophoretic drift that remains constant as this increase in $\Delta D$ compensates for the 
commensurate reduction in diffusivity. 
This effect vanishes for very small particles---such as ions---because $\Delta D \approx 0$ and \cref{eq:ito} is valid for $\alpha = 1$.

 \subsection*{Conclusions}

We designed a microfluidic device platform to observe viscophoretic transport in steep viscosity gradients for the first time. These observations unambiguously prove the existence of this new transport phenomena. Our results show that theoretical predictions break down for nanoparticles larger than at least 9\,nm; they drift substantially faster than expected---likely because predictions do not consider a non-vanishing diffusivity difference across the particles' diameter.  
The existence and understanding of viscophoresis provides a new tool for understanding and controlling particle transport in a variety of contexts such as: the role of viscosity gradient across cell membranes in the uptake of nanoparticles \citep{Behzadi2017}; how biological systems utilize microviscosities to affect molecular trans-membrane transport in organelles \citep{Shalliker2007}, 
and overcoming the main bottleneck to high-performance (solid-state) nanopore sensors \citep{Xue2020, Branton2008, Dekker2007} by using viscosity gradients to slow analyte translocation \citep{Qiu2019}.
Furthermore, the unexpected observation of the accumulation \citep{Mathwig2024} of particles at the intersection of opposing viscophoretic and diffusiophoretic velocity magnitudes is comparable to state-of-the-art molecular traps  such as electrophoretic \citep{Cohen2005}, electrostatic \citep{Krishnan2010}, and entropic traps \citep{Han2000}, optical and magnetic tweezers \citep{Neuman2008}, and nanoflasks \citep{Zhao2016}. 
However, this new \emph{viscophoretic trap} promises three important advantages: 1) no feedback control is necessary, 
2) molecules can be trapped under any electrolyte conditions independent from charge and dielectric properties 
and 3) straightforward up-scaling to parallel trap arrays enabling statistical analysis.

\section*{Methods}

\subsection*{Materials}
Water in all used samples was purified using a MilliQ system (Millipore, Billerica, MA, USA). Carboxylated polystyrene nanoparticles with 660\,nm/680\,nm and 505\,nm/515\,nm excitation/emission wavelength and average diameter of 28\,nm and 110\,nm, respectively, were purchased from Thermofisher scientific (Eugene, Oregon, USA), catalog numbers F8803 and F8783. Carboxylated CdTe quantum dots with a nominal diameter of 8.6\,nm and 780\,nm emission wavelength and excitation wavelength below 400\,nm were purchased from PlasmaChem GmbH (Berlin, Germany). Succinic anhydride, (3-aminopropyl)triethoxysilane (APTES), absolute ethanol, and dimethylformamide (DMF) were purchased from Sigma-Aldrich (The Netherlands). Sylgard 184 polydimethylsiloxane (PDMS) was purchased from Dow Corning (USA) and SU8 photoresist from MicroChem (Newton, MA, USA). Aqueous solutions with glycerol concentrations ranging from 0\% to 50\% (w:w\%) were prepared directly before the measurements. 10\,mM NaOH was added as background electrolyte. After adding fluorescent particles, solutions were vortexed for 2\,min.

\subsection*{Device fabrication, microfluidics, and surface modification}
Stepwise parallel fabrication  consisting of photolithography, soft lithography, and bonding was employed to fabricate microfluidic devices. First, a master mold was fabricated using photolithography in two layers of \SI{2}{\micro\meter} and \SI{200}{\micro\meter} thick SU8-2002 and SU8-2100 negative photoresists, respectively, using spin-coating on 10-cm silicon wafer substrates, soft-baking, UV exposure, development, and post baking steps \citep{Mata2006, Campo2007, Anhoj2007}. Afterwards, the structure was transferred from the fabricated master mold into PDMS to form microchannels. In this step, the ratio of PDMS elastomer to curing agent was kept at 10:1 (w:w) to obtain a high Young’s modulus \citep{Gokaltun2017};  the curing process was carried out at 70\textdegree C for 2\,h on a leveled hotplate. The molded PDMS compartments were subsequently bonded to borosilicate coverslips with \SI{170}{\micro\meter} thickness (24\,mm\,×\,60\,mm, Engelbrecht) following surface activation with oxygen plasma in a Harrick plasma cleaner at 42\,kPa for 30\,s.

The microfluidic device has two inlets channels (and an outlet) which  are \SI{200}{\micro\meter} high, and an array of microchannels with a \SI{2}{\micro\meter}\,×\,\SI{5}{\micro\meter} cross section and different nominal lengths ranging from \SI{100}{\micro\meter} to \SI{150}{\micro\meter} in different device designs. A microchannel array connects both inlet channels. The bypass channel positioned downstream of this array has a \SI{200}{\micro\meter}\,×\,\SI{500}{\micro\meter} cross section and the same length as the channels in the array. A residual pressure difference between both inlets is released via the bypass. Inlet channels were connected to two \SI{500}{\micro\meter} glass syringes (ILS GmbH, Germany) via Tygon tubing with 0.25\,mm/0.76\,mm inner/outer diameter (Avantor VWR, USA). Syringes were filled with liquids of different viscosities; and flow was driven by two Harvard Apparatus 11 Pico Plus syringe pumps set to constant pump rates to generate a stable viscosity gradient along the observation channels. In order to eliminate residual flow in the observation channel, inflow rates were set equal to the inverse of the viscosity ratios during viscophoresis experiments. Mirofluidic inlet channels in the device connect to a single outlet microchannel downstream of the microchannel array and bypass. This outlet was connected to a reservoir open to atmospheric pressure via polytetrafluoroethylene (PTFE) tubing with 0.4\,mm/0.8\,mm inner/outer diameter---sufficiently wide to easily remove air bubbles trapped in the device.

The hydrophobic surface of  PDMS and glass coverslips was modified to prevent the deposition of carboxylated fluorescent particles inside the microfluidic device. We adopted a strategy to functionalize interior surfaces with stable carboxyl groups which deprotonate at pH 12 (10\,mM NaOH base), thus generating negatively charged surfaces for electrostatic repulsion of particles from the microchannel walls. \citep{Kralj2011, Cash2012, An2007} Surface modification followed three successive steps: 1) activation using oxygen plasma (30\,s) to generate hydroxyl groups  on the surfaces,  2) introducing 5\% (v:v) APTES in absolute ethanol via the syringe pump at atmospheric pressure for 1\,h at 60\textdegree C to passivate amine linkers,  and 3) introducing 20\,mM succinic anhydride in DMF at atmospheric pressure for 2\,h at 100\textdegree C to generate carboxyl groups. Then, the treated devices were filled with deionized water and stored at room temperature until experimentation. 

\hspace{-0.3mm}

\subsection*{Microscopy and image analysis}
Fluorescent microscopy was conducted using a home-build setup, openFrame. \citep{openframe} The setup consists of a multimode laser unit with 402\,nm, 520\,nm, and 680\,nm wavelengths (MatchBox, Integrated Optics, Lithuania) operating at 200\,mW, 140\,mW and 360\,mW, respectively. The laser output was coupled into a \SI{150}{\micro\meter}\,×\,\SI{150}{\micro\meter} square silica core, which promotes mode mixing in the fiber in order to obtain a uniform spatial distribution and a square beam shape compared to a Gaussian profile from a single mode fiber for the excitation beam (Thorlabs M102L05). An oil immersion objective (Apo TIRF 60x 1.49 NA, Nikon), a double band emission filter (ZET 532/640m-TRF), and a camera (Prime 95B \mbox{sCMOS}, 202, Photometrics) were implemented. The setup is controlled by the micromanager 1.4 software package, \citep{Edelstein2010} while  the frame time is externally controlled in accordance with the laser pulses using a home build program, SMILE. \citep{smile} Effective pixel sizes of 180\,nm and 360\,nm result from 1\,x\,1 and 2\,x\,2 binning. Camera images are cropped into the microchannels before data acquisition to reach high acquisition rates of 200\,fps and 500\,fps, respectively.

Diffusometry and velocimetry were conducted based on pair Correlation Function (pCF) techniques. \citep{DiRienzo2016, DiRienzo2013,Khandan2024} Image stacks of 68k frames were used to calculate the pCF for each pixel relative to 25 neighboring pixels (for 9\,nm and 28\,nm particles) and 15 neighboring pixels (for 110\,nm particles) neighboring pixels using Fast Fourier Transform for multiple lag times equivalent to 1 up to 12 frames. Then, eight coupled 1D Gaussian distributions were fitted into each calculated pCF at each lag time to quantify the center’s position and Mean Square Displacement (MSD) in eight directions sectioning a 2D surface of 360\textdegree\ into eight angular sectors of 90\textdegree , with 45\textdegree\ overlap between two successive sectors. Afterwards, the drift velocity was calculated using a displacement taken at the center of the pCF. The diffusion coefficient in each direction was calculated using the time evolution of the quantified MSD.

 \section*{Acknowledgment}
 We thank P.\,P.\,M.F.\,A. Mulder for technical support.


\section*{Funding}
This work is part of the research program LocalBioFood with project number 731.017.204 which is financed by the Dutch Research Council (NWO). This work was supported by a Ph.\,D. fellowship (M. F.) from the Graduate School Experimental Plant Sciences to J. H.\,\,\,K. M. acknowledges support by Provincie Gelderland.






\section*{References}
\bibliography{apssamp2}

\end{document}